\documentclass[12pt]{article}
\usepackage{amsmath,amssymb}
\newcommand  {\Ebar} {{\mbox{\rm$\mbox{I}\!\mbox{E}$}}}

\newsavebox{\zzzbar}
\sbox{\zzzbar}
  {\setlength{\unitlength}{0.9em}
  \begin{picture}(0.6,0.7)
  \thinlines
 \put(0,0){\line(1,0){0.6}}
  \put(0,0.75){\line(1,0){0.575}}
  \multiput(0,0)(0.0125,0.025){30}{\rule{0.3pt}{0.3pt}}
  \multiput(0.2,0)(0.0125,0.025){30}{\rule{0.3pt}{0.3pt}}
  \put(0,0.75){\line(0,-1){0.15}}
  \put(0.015,0.75){\line(0,-1){0.1}}
  \put(0.03,0.75){\line(0,-1){0.075}}
  \put(0.045,0.75){\line(0,-1){0.05}}
  \put(0.05,0.75){\line(0,-1){0.025}}
  \put(0.6,0){\line(0,1){0.15}}
  \put(0.585,0){\line(0,1){0.1}}
  \put(0.57,0){\line(0,1){0.075}}
  \put(0.555,0){\line(0,1){0.05}}
  \put(0.55,0){\line(0,1){0.025}}
  \end{picture}}
\newcommand{\Zbar}{\mathord{\!{\usebox{\zzzbar}}}}

\newsavebox{\uuunit}
\sbox{\uuunit}
    {\setlength{\unitlength}{0.825em}
     \begin{picture}(0.6,0.7)
        \thinlines
       \put(0,0){\line(1,0){0.5}}
        \put(0.15,0){\line(0,1){0.7}}
        \put(0.35,0){\line(0,1){0.8}}
       \multiput(0.3,0.8)(-0.04,-0.02){12}{\rule{0.5pt}{0.5pt}}
     \end {picture}}

\newcommand{\QED}{{\hspace*{\fill}\rule{2mm}{2mm}\linebreak}}

\newcommand{\Z}{\Zbar}

\newcommand{\E}{\Ebar}
\newcommand{\p}{\mathbb{P}}

\begin{document}
\setlength{\textheight}{21cm}
\title{{\bf  NO CURRENT WITHOUT HEAT
}}
\author{Christian Maes\thanks{Instituut voor
    Theoretische Fysica, K.U. Leuven, Celestijnenlaan 200D, B-3001
    Leuven, Belgium - email:
    {\tt Christian.Maes@fys.kuleuven.ac.be } },
Frank Redig\thanks{T.U.Eindhoven. On leave from Instituut
    voor Theoretische Fysica, K.U. Leuven, Celestijnenlaan 200D, B-3001
    Leuven, Belgium - email: {\tt f.h.j.redig@tue.nl } } and
Michel Verschuere\thanks{ Instituut voor
    Theoretische Fysica, K.U. Leuven, Celestijnenlaan 200D, B-3001
    Leuven, Belgium - email:
    {\tt Michel.Verschuere@fys.kuleuven.ac.be } }
    }
\maketitle

\footnotesize
\begin{quote}
{\bf Abstract:} We show for a large class of interacting particle systems
that whenever the stationary measure is not reversible for the dynamics,
then the mean entropy production in the steady state is strictly positive.
This extends to the thermodynamic limit the equivalence between
microscopic reversibility and zero mean entropy production: time-reversal
invariance cannot be spontaneously broken. \vspace{3pt}

 \vspace{3pt}

\end{quote}
\normalsize \vspace{12pt}
\renewcommand{\baselinestretch}{2}\normalsize

\section{Introduction}
Reversibility and entropy are words with many meanings even within the
context of nonequilibrium statistical mechanics. One class of models that
has often been considered for learning about nonequilibrium behavior is
that of interacting particle systems. These are stochastic dynamics for
spatially extended systems in which particles locally interact.  They are
mostly toy-models remaining far from realistic in their microscopic
details. Yet, it is believed
 that for some good purposes, the details do not
matter so much and one should be concerned more with the symmetries,
possible conservation laws, locality of the interaction etc.
to hope to understand something about real nature.\\
This paper is about the relation between time-reversal invariance and the
positivity of entropy production.  We do this in the context of
interacting particle systems following the work in
\cite{LS,MRV,M,MRVM,MR}. The physics background will be discussed in
Section 3. The main question is to understand why there cannot exist a
`superconducting' interacting particle system in the sense of the title
of this paper to be specified below.  To understand the mathematical
problem, let us look first at a finite Markov chain. Suppose that $K$ is
a finite set on which we have an involution $\pi: K\rightarrow K,
\pi^2=id$, called time-reversal.  Often, the most natural choice for
interacting particle systems is $\pi = id$ because we think of the state
space as consisting of occupation variables or of classical spins but our
mathematical set-up will be more general.\\ Let $(X_t, t\in[-T,T])$ be a
stationary Markov process (steady state) on $K$ with law $\p_\rho$.  The
subscript refers to the unique stationary probability measure $\rho$ on
$K$; we assume that $\rho(a) > 0, a\in K $.
The rate to go from $a$ to $b$  is denoted by $k(a,b), a,b \in K$
 and we assume that $k(\pi b,\pi a)=0$
iff $k(a,b)=0$ (dynamic reversibility). The generator is
\begin{equation}\label{gen}
Lf(a) = \sum_b k(a,b) [f(b) - f(a)]
\end{equation}
The time-reversed process of $(X_t)$ is the stationary Markov process
$(Y_t, t\in[-T,T])$ on $K$ with $Y_t\equiv \pi X_{-t}$ having transition
rates
\begin{equation}\label{timerevrates}
\tilde{k}(a,b)\equiv k(\pi b,\pi a)\frac{\rho(\pi b)}{\rho(\pi a)}
\end{equation}
 We denote
its law  by $\tilde{\p}_{\rho\pi}$ ($\rho\pi$ is stationary for
$(Y_t)$).  Of course, it easily happens that $\rho=\rho\pi$ and yet,
$\p_{\rho} \neq \tilde{\p}_{\rho\pi}$.   The corresponding generator for
the time-reversed process is $\tilde{L}= \pi L^\ast \pi$ where the $\ast$
refers to the adjoint with respect to the stationary measure
$\rho$.  

We say that the process $(X_t)^T_{-T}$ is {\bf $\pi$-reversible} if
$\p_\rho=\tilde{\p}_{\rho\pi}$.  This implies that the stationary measure
$\rho$ satisfies $\rho=\rho\pi$ and
\begin{equation}\label{db}
\rho(a)k(a,b)=k(\pi b,\pi a)\rho(b), a,b\in K
\end{equation}
which is generalized (or extended) detailed balance (microscopic
reversibility). For the generators, we then have $\tilde{L}=L$.
Observe that (\ref{db}) by itself implies that $\rho(a) = \rho\pi(a)
\rho(b)/\rho\pi(b)$ whenever $k(a,b)\neq 0$.  Applying this successively
with $b_1,\ldots,b_n\in K$ for which
$k(a,b_1),k(b_1,b_2),\ldots,k(b_n,\pi a) \neq 0$, we find that
$\rho(a)=\rho\pi(a)$. On the other hand, $\pi$-reversibility implies that
$\rho=\rho\pi$
is stationary. \\
The entropy production is the random variable obtained from taking the
relative action on pathspace with respect to time-reversal, see \cite{mc}
for a recent review.  Let $\rho_{-T}$ be a probability measure on $K$
which we use to sample the initial data at time $-T$ for the stochastic
time-evolution generated by $L$. The law of this process is denoted by
$\p_{\rho_{-T}}$.  Suppose now that for this process the state at time $T$
is described by the probability measure $\rho_T$.  We could as well start
our process (at time $-T$) from $\rho_T\pi$ and then obtain the process
$\p_{\rho_{T}\pi}$.  For a particular realization $\omega= (\omega(t),t\in
[-T,T])$ of this process we let $\Theta_\pi \omega \equiv (\pi
\omega(-t), t\in [-T,T])$ be its time-reversal. The entropy production $
R_\pi(L,\rho_{-T},T)$ is a function of the realization over the
time-interval $[-T,T]$ and is then obtained as
\begin{equation}\label{enpro}
R_\pi (L,\rho_{-T},T) (\omega)
 = \log \frac{d\p_{\rho_{-T}}} {d\p_{\rho_T\pi}\Theta_\pi}(\omega)
\end{equation}
Here we are only interested in its steady state expectation value, that
is the mean entropy production rate, which in fact can be written as
  \begin{equation}\label{mepfin}
 \textrm{MEP}_{\pi}(L,\rho)=\lim_{T\uparrow\infty}\frac{1}{2T}
 \Ebar_\rho[\log\frac{d\p_\rho}{d\tilde{\p}_{\rho\pi}}]
 \end{equation}
 where $\Ebar_\rho$ denotes expectation
with respect to $\p_\rho$.
 The notation
 MEP$_\pi(L,\rho)$ reminds us that this
number depends on the transformation $\pi$, the dynamics
 (generated via $L$) and the stationary measure $\rho$.  The mean entropy
 production thus measures the degree to which $\p_\rho$ can be
 distinguished from $\tilde{\p}_{\rho\pi}$.
 The  main property of the mean entropy production is then:\\

\noindent {\bf Proposition 1:}  {\it Consider the stationary process}
$(X_t)$ {\it above with}  $\rho=\rho\pi$. Then,
$\textrm{MEP}_{\pi}(L,\rho)=\textrm{MEP}_{\pi}(\tilde{L},\rho)\geq 0$
{\it with equality if and only if the process} $(X_t)$ {\it is}
$\pi$-{\it reversible}.

This says that for finite state space Markov chains
there can be no current without heat, meaning that detailed balance is
equivalent with zero mean entropy production.  The problem we address
here is whether the same remains true in the thermodynamic limit, that is
for spatially extended interacting particle systems.   In this case we
really should be speaking about the mean entropy production {\it
density}, i.e., per unit volume, but we will not use this extension. Note
that in this case and from now on we will not and we
cannot assume in general that $\rho=\rho\pi$ even if both are stationary.\\
 We discuss the general physics set-up and further interpretations in
Section 3, after stating our mathematical results in Section 2.  We start
however with three examples illustrating some aspects.

\subsection{Examples}
{\it Example A:} We consider particles hopping on the one-dimensional
lattice with a preferred direction that is itself subject to independent
flips.  The state space is $\{-1,+1\}\times\{0,1\}^{\Z}$ and the process
is
 determined by choosing a constant rate $c(E,\eta)=1$
  for changes from a configuration $(E,\eta)$ to $(-E,\eta)$ and
taking rates
\[
c(x,E,\eta)= e^E \eta_x(1-\eta_{x+1}) + e^{-E} \eta_{x+1}(1-\eta_x)
\]
for changes to $(E,\eta^{x,x+1})$ where $(\eta^{x,x+1})_y = \eta_y$ if
$x\neq y\neq x+1$, and $(\eta^{x,x+1})_y = \eta_x$ when $y=x+1$ and $=
\eta_{x+1}$ when $y=x$.  The resulting Markov process has generator
\begin{eqnarray}
Lf(E,\eta) &=& \sum_{x} [e^E \eta_x(1-\eta_{x+1}) + e^{-E}
\eta_{x+1}(1-\eta_x)]
[f(E,\eta^{x,x+1})-f(E,\eta)]\nonumber \\
&+& f(-E,\eta)-f(E,\eta)
\end{eqnarray}
For invariant measure $\rho$ we take
\[
\rho(E,d\eta) \equiv \frac 1{2}(\delta_{E,+1} +
\delta_{E,-1})\times\nu_u(d\eta)
\]
where $\nu_u$ is the Bernoulli measure with specified density $u\in
(0,1)$. For
time-reversal we take $\pi(E,\eta)=(-E,\eta)$ so that $\rho=\rho\pi$.\\
It is easy to see that the process satisfies generalized detailed balance,
like (\ref{db}), in the sense that
$\rho(E,d\eta)=\rho(-E,d\eta)=\rho(E,d\eta^{x,x+1})$ and both
\[
c(E,\eta)= c(-E,\eta) \mbox{ and } c(x,E,\eta)= c(x,-E,\eta^{x,x+1})
\]
The last identity depends of course crucially on the fact that $\pi$ is
not the identity and reverses left and right as preferred direction.  At
the same time, as can be computed explicitly, the mean entropy production
is zero.  The same remains true for $\pi$ a particle-hole transformation,
$(\pi\eta)_x=1-\eta_x$, leaving the field $E$ unchanged.
  Then, $\rho\neq \rho\pi$ for $u\neq 1/2$
but still generalized detailed balance holds.  Finally if, instead, we
were to take $\pi=$ identity as time-reversal, then we break the detailed
balance condition and we obtain a strictly positive mean entropy
production.

\noindent {\it Example B:} We take the simplest example of a spinflip
dynamics for which the one-dimensional Ising model is stationary but not
reversible (for $\pi=$id).  Exactly the same can be done in two
dimensions, see \cite{Ku}.  Spinflips are transformations
$U_x:\sigma\rightarrow U_x(\sigma)= \sigma^x, x\in \Z, \sigma \in
 \{+1,-1\}^{\Z}$ for $\sigma^x$ equal to $\sigma$ except at the site
 $x$.\\
Consider the one-dimensional spinflip dynamics with the following
asymmetric rates:
\begin{equation}\label{ku}
 c(x,\sigma)= \exp (-2\beta \sigma_x\sigma_{x+1} )
  \end{equation}
   The invariant measure
$\rho$ is the one-dimensional Ising model at inverse temperature $\beta$.
The process starting from $\rho$ is not time-reversal invariant and the
entropy production is equal to MEP$(L,\rho) = 4\beta\tanh \beta$ (that is
with time-reversal $\pi=$identity). On the other hand, this time-reversed
process is easy to find; it is a spinflip process with generator
\[
L^\ast f(\sigma) = \sum_x e^{-2\beta\sigma_x \sigma_{x-1}}
[f(\sigma^x)-f(\sigma)]
\]
Let us now take for time-reversal $\pi$ the reflection:
$(\pi\sigma)_x=\sigma_{-x}$ which leaves $\rho$ invariant.  Since
\[
(\pi\sigma)^x = \pi(\sigma^{-x})
\]
$L^\ast = \pi L \pi$ and we have in fact generalized detailed balance
(\ref{db}):
\[
\frac{c(x,\sigma)}{c(-x,(\pi\sigma)^{-x})} = \frac{d\rho\circ
U_x}{d\rho}(\sigma)= e^{-2\beta\sigma_x(\sigma_{x-1} + \sigma_{x+1})}
\]
The denominator in the left hand side is the rate in the original process
by which $\pi U_x \sigma = \pi(\sigma^x)= (\pi\sigma)^{-x}$ is changed to
$\pi\sigma$.
As a result, MEP$_\pi(L,\rho)=$MEP$_\pi(L^\ast,\rho)=0$.

\noindent {\it Example C:} Instead of driving the system in the bulk and
breaking detailed balance via some external fields that act on each
component of the system, we may also consider boundary driven processes.
For this we need to start with finite volume.  The simplest interesting
case is that of a symmetric exclusion process on a lattice interval that
is driven by independent birth and death processes at its boundaries
corresponding to different chemical potentials.  Take
$\Lambda_n=\{-n,-n+1,\ldots,n-1,n\}$ and $\eta\in \{0,1\}^{\Lambda_n}$ a
particle configuration evolving with generator
\begin{eqnarray}\label{gn}
G_n f(\eta) &=& \sum_{x=-n}^{n-1} [f(\eta^{x,x+1})-f(\eta)] \nonumber \\
&+& \lambda[ e^{h_1\eta_{-n}}(f(\eta^{-n}) - f(\eta)) +
e^{h_2\eta_n}(f(\eta^n)-f(\eta))]
\end{eqnarray}
The first term corresponds to symmetric hopping with exclusion; the two
last terms are giving birth and death to particles at the ends of the
interval with parameters $h_1,h_2$.  One can think here of particle
reservoirs, to the left of the
 system with density $1/(1+e^{h_1})$ and
to the right with density $1/(1+e^{h_2})$.  For $\lambda=0$ the system is
uncoupled from the reservoirs and it has all uniform product measures as
reversible measures with vanishing mean entropy production.  For
$\lambda\neq 0, h_1\neq h_2$ this detailed balance
 is lost and we have positive mean entropy production.
Yet, it remains of order unity, uniformly in the size $n$ meaning that the
mean entropy production density vanishes in the thermodynamic limit. This
is an instance of a more general fact for interacting particle systems
that will also be treated in the next section: you cannot by driving the
system at its boundaries break the time-reversal invariance in the
limiting infinite volume process, see Proposition 2 below.\\ We do not
know whether there exists a time-reversal $\pi$ for which (\ref{gn})
would give rise to generalized detailed balance.

In this paper we show more generally how  breaking of detailed balance is
strictly equivalent with non-zero mean entropy production. There is no
way to get a current and at the same time to have no dissipation (zero
mean entropy production).

In the next section we describe our class of models and we state our main
result. In section 3, we discuss this result and we give some more
background information concerning entropy production, reversibility and
time-reversal.  Section 4 is devoted to the proofs.
\section{Models and main result}
\subsection{Dynamics}\label{confi}
This subsection describes the assumptions and introduces the necessary
notation.\\
The configuration space is $\Omega\equiv S^{\Z^d}$ where $S$ is a finite
set and $\Z^d$ is the regular $d-$dimensional lattice.
 Let $\pi$ be an
involution  on $\Omega$. A special but important case is when
$\pi=$identity. We assume here that $\pi$ commutes with lattice
translations
$\tau_x, x\in \Z^d$.\\
  Let $V_0\subset\Z^d$ be a finite cube
containing the origin and write $\mathcal{P}_0$ for any specific
non-empty set of transformations $U_0$ on $\Omega$ satisfying, for every
$U_0\in\mathcal{P}_0$, and for every $\sigma\in \Omega$: \makeatletter
\renewcommand{\theenumi}{\roman{enumi}}
\renewcommand{\labelenumi}{\theenumi.}
\makeatother
\begin{enumerate}\label{u3}
\item $\left(U_0\sigma\right)(y)=\sigma(y)$,
for $y\in V^c_0$,
\item
$U_0^{-1}\in \mathcal{P}_0$,
\item $\pi \mathcal{P}_0 \pi = \mathcal{P}_0$,
\item If $U_0\not= U_0'$ and $U_0 \sigma \not=\sigma, U_0'\sigma  \not=\sigma$
then
$U_0\sigma \not= U'_0\sigma$ (for convenience only.)
\end{enumerate}
We consider the translations $V_x\equiv \{ y+x:y\in V_0 \}$ and $U_x
\equiv \tau_x U_0 \tau_{-x}$ to generate a dynamics via local translation
invariant rates $c(U_x,\sigma)$ for the transition
$\sigma\to\,U_x\sigma$.  We assume:
\begin{enumerate}
\setcounter{enumi}{4}
\item Positivity: $c(U_0,\sigma)=0$ when $U_0\sigma=\sigma$ and if not,
$c(U_0,\sigma) > 0$,
\item Finite range: there is a finite $\bar{\Lambda}\subset\Z^d$
such that for all $\sigma,\eta\in \Omega$, and $U_0\in\mathcal{P}_0$:
$c(U_0,\sigma)=c(U_0,\sigma_{\bar{\Lambda}}\eta_{\bar{\Lambda}^c})$,
\item Translation invariance: for all $x\in\Z^d$, $U_x\in\mathcal{P}_x$,
$\sigma\in \Omega$: $c(U_x,\sigma)=c(U_0,\tau_{-x}\sigma)$
\end{enumerate}
The generator $L$ corresponding to the given rates is now defined on
local functions $f$ as
\begin{equation}\label{GEN}
Lf(\sigma) \equiv
\sum_{x\in\Z^d}\,\sum_{U_x\in\mathcal{P}_x}c(U_x,\sigma)[f(U_x\sigma)
-f(\sigma)]
\end{equation}
That is, $\sigma$ is changed to $\eta$ at rate $c(U_x,\sigma)$ if
$\eta=U_x\sigma$. We will always write $\rho$ for a translation invariant
stationary measure for this dynamics. It can be different from $\rho\pi$
but we assume that also $\rho\pi$ is stationary. Finally, $\rho$ and
$\rho\pi$ give positive weight to all cylinders and writing $\rho^{U}=
U\rho$, we always assume that $d\rho^{U_0}/d\rho(\sigma)\geq c>0$, which,
even in the present rather general set-up, can be expected quite
generally.

For $V_0=\{0\}$ and $S=\{+1,-1\}$, the choice $U_x\sigma=\sigma^x$
corresponds to a spinflip process. Taking $V_0=\{0,e_1,e_2,\ldots,e_d\}$
with $e_\alpha$ the lattice unit vectors, we can make a spin exchange
process or hopping dynamics. We refer to \cite{LIGG} for further details
on constructing the infinite volume process.
\subsection{Mean Entropy Production}
Put $\Lambda_n=[-n,n]^d\cap\Z^d$ for large $n$ and define
$\Lambda^\sharp_n$ as the maximal subset of $\Lambda_n$, such that for all
$x\in\Lambda^\sharp_n$ and $U_x\in\mathcal{P}_x$, $c(U_x,\sigma)$ depends
only on coordinates inside $\Lambda_n$, and $V_x\subset\Lambda$.
 Consider now
the Markov chain on $S^{\Lambda_n}$ with generator
\begin{equation}\label{ln}
L_n f(\sigma)\equiv \sum_{x\in\Lambda^\sharp_n}\sum_{U_x\in\mathcal{P}_x}
 c(U_x,\sigma)[f(U_x\sigma)-f(\sigma)]
 \end{equation}
and started from a probability measure $\rho_{-T}$ on $S^{\Lambda_n}$ at
time $-T$. The measure at time $T$ is denoted by $\rho_T$. Via a Girsanov
formula this dynamics gives rise to a Hamiltonian (or action functional)
on space-time trajectories $\omega$ (as in \cite{mc,MRV}),
 with corresponding relative energy
with respect to time-reversal given by the entropy production
(\ref{enpro}) and here equal to
\begin{equation}\label{entropyproduction}
R_\pi(L_n,\rho_{-T},T)(\omega)= \ln \rho_{-T}(\omega(-T)) -\ln
\rho_T(\omega(T)) + \Delta S_e (\omega)
\end{equation}
with
\begin{eqnarray}\label{vrg}
\Delta S_e(\omega) &=&
 \sum_{x\in\Lambda^\sharp_n}
\sum_{U_x\in\mathcal{P}_x} \int_{-T}^T\log\frac{c(U_x,\omega(s^-))}
{c(\pi U^{-1}_x\pi, \pi
U_x\omega(s^-))}dN^{U_x}_s(\omega)\nonumber\\&+&\int^T_{-T}[c(U_x,\pi\omega(s))-
c(U_x,\omega(s))]ds
\end{eqnarray}
where $N^{U_x}_t(\omega) \equiv \sum_{-T\leq s\leq t} I\left(\omega (s) =
U_x(\omega(s^-))\not=\omega(s^-)\right)$ is the number of times the
transformation $U_x$ appeared in the realization $\omega$ up to time
$t\in [-T,T]$.  The expression (\ref{vrg}) must be interpreted as the
variable entropy produced in the reservoirs (environment) when the
microscopic system configuration moves from $\omega(-T)$ to $\omega(T)$:
To get the total variable entropy production (\ref{entropyproduction})
one should add to (\ref{vrg}) the corresponding change in the system's
entropy, that are the first two
 terms in (\ref{entropyproduction}).
However, when taking steady state averages, this part vanishes (the
entropy of the stationary system does not change on average). We can
therefore define the mean entropy production for the interacting particle
system as
\begin{equation}\label{a}
\textrm{MEP}_\pi(L,\rho)\equiv
\lim_n\lim_{T\uparrow\infty}\frac{1}{2|\Lambda_n|T} \E^{n,T}_\rho(\Delta
S_e)
\end{equation}
$\E^{n,T}_{\rho}$ denotes the expectation with respect to the path space
measure, in the stationary distribution $\rho$, restricted to trajectories
within $S^{\Lambda_n}$. In other words, the mean entropy production is
the expectation of the time-reversal breaking part in the space-time
action functional governing the dynamics.  We refer to \cite{MRV} for a
mathematical discussion on the existence of the limit (\ref{a}) and for a
proof of its non-negativity. We refer to \cite{M,MRVM,mc} and Section 3
for further background.

\subsection{Results}
The main question is to see whether for a dynamics where the time-reversal
symmetry is explictly broken (in the sense that there is no detailed
balance), there still can be zero mean entropy production
(dissipationless steady state).  Our main result says that this is
impossible.

\noindent {\bf Main Theorem:} {\it Under the conditions above,
$\mbox{MEP}_\pi(L,\rho)= \mbox{MEP}_\pi(L,\rho\pi) = 0$ implies that the
dynamics  satisfies (generalized) detailed balance in the sense that for
all} $U_0$
\begin{equation}\label{gendb}
c(\pi U^{-1}_0\pi,\pi
U_0\sigma)\frac{d\rho^{U_0}}{d\rho}(\sigma)=c(U_0,\sigma)\quad\rho- a.s.
\end{equation}
Note that (\ref{gendb}) is really the analogue of (\ref{db}).
Observe also here that (\ref{gendb}) implies that the densities
$d\rho^{U_0}/d\rho$ are invariant under replacing $\rho$ by $\rho\pi$.
This follows from rewriting (\ref{gendb}) from right to left with $\sigma
\rightarrow \pi U_0\sigma$ and $U_0 \rightarrow \pi U^{-1}_0 \pi$:
\begin{eqnarray}\nonumber
c(\pi U^{-1}_0 \pi,\pi U_0\sigma) &=& c(\pi\pi U_0\pi\pi,\pi\pi
U^{-1}_0\pi\pi U_0\sigma) \frac{d\rho^{\pi U^{-1}_0\pi}}{d\rho}(\pi
U_0\sigma) \\\nonumber & = & c(U_0,\sigma)
\frac{d\rho\pi}{d\rho\pi^{U_0}}(\sigma)
\end{eqnarray}
and comparing it with the original (\ref{gendb}).

We call $\mathcal{P}_0$ complete if every local transformation $h:
\Omega\to\Omega$
can be written as
a composition of $U_x$: i.e.,
if
$h=U_{x_1}\ldots U_{x_n}$ for some $x_1,\ldots, x_n \in \Z^d$.

\noindent {\bf Corollary 1:}  {\it If}  $\mbox{MEP}_\pi(L,\rho) =
\mbox{MEP}_\pi(L,\rho\pi) =0$ {\it  and if } $\mathcal{P}_0$ {\it is complete and} $\pi$
 {\it is continuous, then}
$\rho$ {\it is a reversible Gibbs measure for the dynamics defined above}.

In \cite{MRV} the converse to these results was already shown: Suppose
that the rates satisfy
\begin{equation}\label{detbal1}
c(U_x,\sigma )= c(\pi U^{-1}_x \pi, \pi U_x \sigma ) \exp (-H(U_x \sigma )
+ H(\sigma)).
\end{equation}
This is again the analogue of (\ref{db}). The energy difference in
(\ref{detbal1}) should be interpreted in terms of an absolutely convergent
sum of potentials:
\begin{equation}
H(\sigma_\Lambda\eta_{\Lambda^c})- H(\xi_\Lambda\eta_{\Lambda^c}) =
\sum_{A\cap \Lambda\not=\emptyset}
\left(V(A,\sigma_\Lambda\eta_{\Lambda^c})-
V(A,\xi_\Lambda\eta_{\Lambda^c})\right),
\end{equation}
where $(V(A,\cdot):S^A \rightarrow (-\infty,+\infty), A$ finite subsets
of $\Z^d)$, is a translation invariant (uniformly) absolutely summable
potential:
\begin{equation}\label{as}
\sum_{A\ni 0} \max_{\sigma\in S^A} |V(A,\sigma)| < +\infty
\end{equation}
Then,
\[
\rm{MEP}_\pi(L,\rho)=\rm{MEP}_\pi(L,\rho\pi)=0
\]
When we combine the above we obtain a final

\noindent {\bf Corollary 2:}  {\it Under the conditions of Corollary 1,
if there is one translation invariant stationary measure} $\rho$ {\it for
which} $\rho=\rho\pi$ {\it and} MEP$_\pi(L,\rho) =0$, {\it then also}
MEP$_\pi(L,\nu) =0$ {\it for all translation invariant stationary
measures $\nu$ and they are all Gibbsian for the same potential.}

A {\it caveat}
 in the above main result is to understand better the relation between
 MEP$_\pi (L,\rho)$ and MEP$_\pi (L,\rho\pi)$.
To this we can only add that
 MEP$_\pi (\pi L \pi, \rho)=$ MEP$_\pi (L,\rho\pi)$,
as can be verified from a
 direct computation starting with (\ref{result}).
The simplest illustration of all this was already obtained in \cite{MR}
for a spinflip process. Example B, (\ref{ku}), deals with a spinflip
process but there the time-reversal $\pi$ does not commute with
translations.  As will be seen from the proof, that is indeed not
essential as long as the dynamics and the stationary measure are
translation invariant.  Of course, one should then be extra careful with
condition iii. but also this can be modified accordingly.
  It will also be clear that more general lattice
structures and configuration spaces can be employed (e.g. already in Example A).\\
Finally, for completeness we come back to the situation of Example C in
Section 1.1.  For this we must leave the translation invariant infinite
volume context and ask whether boundary driven interacting particle
systems can give rise to non-vanishing mean entropy production density in
the thermodynamic limit.  The question can be formalized as follows.  We
consider a process on $S^{\Lambda_n}$ with generator $G_n$ generalizing
(\ref{gn})
\[
G_n f(\sigma)\equiv L_nf(\sigma) +
\sum_{\begin{array}{c}A\subset\Lambda_n\setminus\Lambda^{\sharp}_n\\
\mbox{\tt diam} A \leq r\end{array}}\sum_{\eta\in
S^A}k^{(n)}_A(\sigma,\eta)[f(\sigma^{A,\eta})-f(\sigma)]
\]
where $\sigma^{A,\eta}\equiv\sigma_{A^c}\eta_A$  equals $\sigma$ outside
the set $A$ which has a diameter (maximal lattice
distance within) less than a given constant $r$.\\
Here the generator $L_n$ is given by (\ref{ln}) but with rates verifying
condition (\ref{detbal1}) for a finite range potential, and rates
$k^{(n)}_A(\sigma,\eta)$ as in (\ref{gn}) inducing configurational
changes at the boundary of $\Lambda_n$.  We further assume that the
$k^{(n)}_A(\sigma,\eta)$ are uniformly bounded from below and from above.
In other words, we have a bulk dynamics generated by $L_n$ with rates
satisfying (generalized) detailed balance, and at the boundary the
configuration can change quite arbitrarily (but in a local and bounded
way). We suppose that $\rho_n$ is the unique stationary measure of this
dynamics and for simplicity we only treat the case $\pi=id$. We are
interested in the mean entropy production MEP$(G_n,\rho_n)$ defined in
(\ref{mepfin}) (with $\pi=id$). \vspace{3pt}

\noindent{\bf Proposition 2:} {\it There is a constant $K$ so that}
MEP$_\pi(G_n,\rho_n) \leq K n^{d-1}$ \vspace{3pt}

The proofs of the above results are postponed to Section 4.
\section{Discussion}
We briefly discuss some concepts that are important for our result.
\subsection{Time-reversal}
By this we usually mean a transformation on phase space $\Omega$ which,
for a many-particle system, is defined particle-wise or, for spatially
extended systems, is sufficiently local. Physically speaking, its precise
nature follows from kinematical considerations on the dynamical
variables.  In classical mechanics, it reverses the momenta of all the
particles but in the presence of say an electromagnetic potential,
considered part of the system,
 one can add an extra transformation
reversing also the magnetic field and thus making the Lorentz force
time-reversal invariant.
  In our case, we have a configuration space $\Omega = S^{\Z^d}$
with $\Z^d$ the $d$-dimensional lattice and $S$ a finite set.
Time-reversal is an involution $\pi$ on $\Omega$, $\pi^2=id$.
Time-reversal extends to a transformation $\Theta_\pi$ on path-space, as
introduced for (\ref{enpro}), by reversing the trajectories.  That is, if
we have a trajectory $(\omega_t, t\in [-T,T])$ then the time-reversed
trajectory $\theta_{\pi}(\omega)$ is given by $(\theta_{\pi}(\omega))_t
\equiv \pi \omega_{-t}$.
\subsection{Reversibility}
Dynamic reversibility is a property of the dynamics itself under
time-reversal.  It says that if one trajectory $\omega$ of the system is
possible, so is its time-reversed $\theta_\pi(\omega)$.  For a
deterministic system where $\omega_t =\phi(t) \omega_0$ with $\phi(t)$ an
invertible flow on phase space, it says that $\phi(t)^{-1} = \pi \phi(t)
\pi$, that is a symmetry that anticommutes with the time evolution.
  For a stochastic dynamics
this is implied by assuming that if a transition $\sigma\rightarrow
U\sigma$ is possible (positive transition rate), then also the same is
true for its time-reversal $\pi U \sigma \rightarrow \pi
\sigma$.\\
Microscopic reversibility is a consequence of dynamic reversibility in
case of an equilibrium dynamics.  For our purposes here we do not make a
distinction with the condition of detailed balance.  When the dynamics is
driven away from equilibrium, the resulting stochastic model will not
satisfy detailed balance.  Usually this produces a current in the system
(but that need not be true in general, see an example in \cite{MRVM}). On
the other hand, a net current signifies the breaking of the detailed
balance condition. In general we like to distinguish between two classes
of finite volume dynamics where microscopic reversibility is explicitly
broken. These are boundary driven versus bulk driven dynamics depending
on the extensivity of the perturbation from an equilibrium dynamics. In
the bulk driven case, one usually verifies so called {\it local detailed
balance}, i.e., (\ref{detbal1}) is changed into
\[
c(U_x,\sigma )= c(\pi U^{-1}_x \pi, \pi U_x \sigma ) \exp (-H(U_x \sigma )
+ H(\sigma)) e^{E \Phi(U_x\sigma,\sigma)}
\]
where $E$ is some amplitude of an external field and $\Phi$ is
antisymmetric, $\Phi(\pi \eta,\pi \sigma)=-\Phi(\sigma,\eta)$, see e.g.
\cite{LS}. Note also that then, necessarily, the relative energies
$H(U_x \sigma) - H(\sigma)$ are invariant under exchanging $H$ with $H\pi$.\\
In boundary driven systems, the process becomes non-translation invariant
and the rates remain of the form (\ref{detbal1}) in the bulk (that is for
$x$ well inside the considered finite volume) while more or less
arbitrary on the boundary.  This was the case for Example C in Section
1.1 and was formalized for Proposition 2.  Note that there is in fact an
example of a boundary driven system where uniformly in the size of the
system a bulk current can be maintained. This is the nonequilibrium
harmonic crystal treated in \cite{RLL,N} where the heat flux is
proportional to the boundary temperature difference rather than to the
temperature gradient (infinite heat conductivity in the thermodynamic
limit). Such `superconductors' do not exist in the context of interacting
particle systems as discussed in the present paper.

\subsection{Entropy production}
In phenomenological thermodynamics, entropy production appears in open
driven systems as the product of thermodynamic fluxes and forces.  The
forces are gradients of intensive quantities (like chemical potential)
generating the currents.  The entropy production is identified from a
balance equation for the time-derivative of an entropy density
 which is defined in systems close to equilibrium.   The definition of
 entropy production as we use it here in statistical mechanics comes from
 \cite{M,MRV,MRVM,MR,Q,QQ,S} and we refer to the review \cite{mc}.
The mean entropy production appears there and in
(\ref{enpro})-(\ref{mepfin}) as a relative entropy (density) for the
process with respect to its time-reversal.  That immediately invites the
following thought (we are grateful to Senya Shlosman for pointing to
this): In equilibrium statistical mechanics, if two translation invariant
Gibbs measures have zero relative entropy density, then they must both be
Gibbsian for the same interaction potential (but not necessarily equal
e.g. because of spontaneous symmetry breaking). Apply this to the
space-time measures obtained for the process $\p_\rho$ and the
time-reversed process $\p_{\rho}\Theta$ as introduced for (\ref{enpro}).
Here we take $\pi=$ identity to avoid extra complications. In some sense,
both processes
 are Gibbs measures.  Thus, if
the mean entropy production is zero, then the process itself and its
time-reversal have the same (space-time) action functional.  Because they
also have the same marginals $\rho$, they must in fact coincide (hence no
spontaneous time-reversal breaking).  Hence, zero mean entropy production
implies microscopic reversibility. While convincing on a superficial
level, unfortunately the details of this argument are technically
cumbersome and a direct sufficiently general proof along this line has
not been found.

The only more recent paper that we know of concerning time-reversal
symmetry and the relation with entropy production is \cite{G}.  The set-up
there is however quite different from ours.  Time-reversal symmetry is
there associated with the anticommutation of an involution with the time
evolution, what we have called dynamic reversibility in the above.  In
our discussions here, we deal with spatially extended stochastic dynamics
and the breaking of microscopic reversibility.

\section{Proofs}
\noindent {\bf Lemma 1:} {\it Under the conditions of Section 2.1, for a
translation invariant stationary measure} $\nu$,
\begin{equation}
\sum_{U_0\in\mathcal{P}_0}\int
d\nu(\sigma)c(U_0,\sigma)\log\frac{d\nu^{U_0}}{d\nu}(\sigma)=0
\end{equation}

\noindent {\bf Proof:} Let $\mathcal{F}_\Lambda$ be the $\sigma$- field
generated by $\sigma_x, x\in\Lambda$.  Denote by $\nu_\Lambda$,
respectively $\nu^{U_0}_\Lambda$ the $\mathcal{F}_\Lambda$- restrictions
of $\nu$ and $\nu^{U_0}$.  Then we have
\begin{equation*}
\frac{d\nu^{U_0}_\Lambda}{d\nu_\Lambda}= \Ebar_\nu
\left[\frac{d\nu^{U_0}}{d\nu}|\mathcal{F}_\Lambda\right]
\end{equation*}
Since $d\nu^{U_0}/d\nu\in L^1(d\nu)$ for all $U_0$, we find using the
martingale convergence theorem that
\begin{equation}
\lim_{\Lambda\uparrow\Z^d}\frac{d\nu^{U_0}_\Lambda}{d\nu_\Lambda}=
\frac{d\nu^{U_0}}{d\nu},
\end{equation}
in $L^1(d\nu)$.  Let $\tilde{\nu}$ be the product measure on $\Omega$
having
 as marginals the uniform measure on $S$. From stationarity applied to the
local function $f_\Lambda=d\nu_\Lambda/d\tilde{\nu}_\Lambda$ we find
\begin{eqnarray*}
0&=&\sum_{x\in\Lambda'}\sum_{U_x\in\mathcal{P}_x}\int
d\nu(\sigma)c(U_x,\sigma)
[\log\frac{d\nu^{U_x}_\Lambda}{d\tilde{\nu}_\Lambda}-
\log\frac{d\nu_\Lambda}{d\tilde{\nu}_\Lambda}]\\
&=&\sum_{x\in\Lambda'}\sum_{U_x\in\mathcal{P}_x}\int
d\nu(\sigma)c(U_x,\sigma)\log\frac{d\nu^{U_x}_\Lambda}{d\nu_\Lambda}\\
&=&\sum_{x\in\Lambda'}\sum_{U_x\in\mathcal{P}_x}\int
d\nu(\sigma)c(U_x,\sigma)\log\frac{d\nu^{U_x}}{d\nu}\\
&+&\sum_{x\in\Lambda'}\sum_{U_x\in\mathcal{P}_x}\int
d\nu(\sigma)c(U_x,\sigma)
\left[\log\frac{d\nu^{U_x}_\Lambda}{d\nu_\Lambda}-
\log\frac{d\nu^{U_x}}{d\nu}\right]\\
&=&|\Lambda'|\sum_{U_0\in\mathcal{P}_0}
\int d\nu(\sigma)c(U_0,\sigma)\log\frac{d\nu^{U_0}}{d\nu}(\sigma)\\
&+&\sum_{x\in\Lambda'}\sum_{U_x\in\mathcal{P}_x}\int
d\nu(\sigma)c(U_x,\sigma)F^{U_x}_\Lambda(\sigma).
\end{eqnarray*}
The last equality uses translation invariance. We have used the notation
$\Lambda' \equiv \{x\in\Z^d|V_x\cap\Lambda\neq\emptyset\}$ and the
expression\[F^{U_x}_\Lambda(\sigma)\equiv
\left(\log\frac{d\nu^{U_x}_\Lambda}{d\nu_\Lambda}-
\log\frac{d\nu^{U_x}}{d\nu}\right)\] We thus have
\begin{eqnarray}
|\sum_{U_0\in\mathcal{P}_0}\int d\nu(\sigma)c(U_0,\sigma)
\log\frac{d\nu^{U_0}}{d\nu}(\sigma)
|&\leq&\frac{1}{|\Lambda'|}\sum_{x\in\Lambda'}\sum_{U_x\in\mathcal{P}_x}|\int
d\nu(\sigma)c(U_x,\sigma)F^{U_x}_\Lambda(\sigma)|\nonumber\\
&\leq&M\frac{1}{|\Lambda'|}
\sum_{x\in\Lambda'}\sum_{U_0\in\mathcal{P}_0}\int
d\nu|F^{U_0}_{\Lambda-x}|,
\end{eqnarray}
by the translation invariance of $\nu$, and $M$ bounds the rates. Now we
use the general fact that if $f_n$ converges to $f$ in $L^1(d\nu)$ and
both $f_n, f$ are bounded from below by some constant $c>0$, then $\log
f_n$ converges to $\log f$ in $L^1(d\nu)$. This fact implies that for any
given $\varepsilon>0$, we can choose $\Delta\subset\Z^d$ such that for
all $\Delta'\supset\Delta$:
\[
\max_{U_0\in\mathcal{P}_0} \int
d\nu|F^{U_0}_{\Delta'}|\leq\frac{\varepsilon}{2M N},\quad\mbox{with}\quad
|\mathcal{P}_0| \equiv N
\]
 Choose now
$\Lambda\subset\Z^d$ so large that
\[
\frac{|\{x\in \Lambda': \Delta+x\cap\Lambda^c \neq
\emptyset\}|}{|\Lambda'|} \leq\frac{\varepsilon}{2M
N\sup_{W,U_0}||F^{U_0}_W||_{L^1(d\nu)}}
\]
We then conclude that
\begin{eqnarray}
|\sum_{U_0\in\mathcal{P}_0} \int d\nu(\sigma)
 & c(U_0,\sigma) &
\log\frac{d\nu^{U_0}}{d\nu}(\sigma) | \leq
\frac{1}{|\Lambda'|}\sum_{x\in\Lambda',\Delta+x\subset\Lambda}
\frac{\epsilon}{2} +\nonumber\\ & M N & \frac{|\{x\in
\Lambda':\Delta+x\cap\Lambda^c \neq \emptyset\}|}{|\Lambda'|}
\sup_{W,U_0}||F^{U_0}_W||_{L^1(d\nu)}\nonumber\\ & \leq & \varepsilon
\end{eqnarray}
\QED \noindent {\bf Proof of Main Theorem:} Define
\[
\bar{c}(U_0,\sigma)\equiv\bar{c}(\pi,\rho;U_0,\sigma)\equiv c(\pi
U^{-1}_0\pi,\pi U_0\sigma)\frac{d\rho^{U_0}}{d\rho}(\sigma)
\]
and substitute it in
\begin{equation}\label{compute}
\sum_{U_0\in\mathcal{P}_0}\int d\rho(\sigma)
[c(U_0,\sigma)-\bar{c}(U_0,\sigma)]
\log\frac{c(U_0,\sigma)}{\bar{c}(U_0,\sigma)}
\end{equation}
We get four terms, (\ref{compute}) =
\begin{eqnarray}\label{fourt}
&& \sum_{U_0\in\mathcal{P}_0}[\int d\rho(\sigma) \log
\frac{c(U_0,\sigma)}{c(\pi\,U_0^{-1}\pi,\pi\,U_0\sigma)} +  \int
d\rho(\sigma) c(U_0,\sigma) \log \frac{d\rho}{d\rho^{U_0}} \\\nonumber +
&& \int d\rho^{U_0}(\sigma) c(\pi U_0^{-1}\pi,\pi U_0\sigma) \log
\frac{c(\pi\,U_0^{-1}\pi,\pi\,U_0\sigma)}{c(U_0,\sigma)} + \int
d\rho^{U_0}(\sigma) c(\pi U_0^{-1}\pi,\pi U_0 \sigma) \log
\frac{d\rho^{U_0}}{d\rho}]
\end{eqnarray}
The second term is zero by Lemma 1.  The fourth term is also zero because,
using condition iii, we can change $\pi U_0^{-1} \pi\rightarrow U_0$ in
the sum over $\mathcal{P}_0$ getting it equal to
\[
\sum_{U_0\in\mathcal{P}_0} \int d\rho\pi(\sigma) c(U_0,\sigma) \log
\frac{d\rho\pi}{d(\rho\pi)^{U_0}}
\]
which is zero, again by Lemma 1 applied to the stationary measure
$\rho\pi$. Again using iii, we can also
 rewrite the third term as
 \[
\sum_{U_0\in\mathcal{P}_0} \int d\rho\pi(\sigma) \log
\frac{c(U_0,\sigma)}{c(\pi\,U_0^{-1}\pi,\pi\,U_0\sigma)}
 \]
Therefore, what remains of (\ref{fourt})
 is the sum of the first and the third term so that (\ref{compute}) equals
 \[
 \sum_{U_0\in\mathcal{P}_0}[\int d\rho(\sigma) \log \frac{c(U_0,\sigma)}{c(\pi\,U_0^{-1}\pi,\pi\,U_0\sigma)}
 + \int d\rho\pi(\sigma) \log
 \frac{c(U_0,\sigma)}{c(\pi\,U_0^{-1}\pi,\pi\,U_0\sigma)}]
 \]
 We now recall that the mean
entropy production (\ref{a}) equals
\begin{eqnarray}\label{result}
\textrm{MEP}_\pi(L,\rho)&=& \sum_{U_0\in\mathcal{P}_0}\, (\int
d\rho(\sigma) \, c(U_0,\sigma)
\log\frac{c(U_0,\sigma)}{c(\pi\,U_0^{-1}\pi,\pi\,U_0\sigma)}
\nonumber\\
&+&\int d\rho(\sigma)\,[c(U_0,\pi\sigma)-c(U_0,\sigma)])
\end{eqnarray}
This was derived from (\ref{vrg}) in \cite{MRV}. We conclude therefore
that (\ref{compute}) equals $\textrm{MEP}_\pi(L,\rho) +
\textrm{MEP}_\pi(L,\rho\pi)$ which is zero by hypothesis.  This implies
the statement of the Theorem.\QED

\noindent {\bf Proof of Corollary 1} Since the Radon-Nikodym derivative of
$\rho^{U_0}$ with respect to $\rho$ is a local function for all $U_0$ and
since by assumption, we can generate with the $U_0$ all local excitations
$\sigma'$ from $\sigma$, it means that $\rho$ has a continuous version
for its local conditional distributions. \QED

\noindent {\bf Proof of Corollary 2} From the main result and Corollary 1
it follows that $\rho$ is a translation invariant stationary Gibbs measure
and (\ref{detbal1}) must be satisfied. All other translation invariant
stationary measures must be Gibbsian and for the same potential, see e.g.
\cite{Ku}.
 From the
results in \cite{MRV} as cited above the statement of Corollary 2, it
follows that every other stationary translation invariant measure must
have zero mean entropy production. \QED

\noindent {\bf Proof of Proposition 2}  From the definition
(\ref{mepfin}) we must first compute the relative action under time
reversal, that is
\[
R_n\equiv\log\frac{d\p_{\rho_n}}{d\tilde{\p}_{\rho_n}}
\]
This can be done via a Girsanov formula and we obtain the analogue of
(\ref{vrg}):
\begin{eqnarray*}
R_n(\omega)&=& \sum_{x\in\Lambda^\sharp_n} \sum_{U_x\in\mathcal{P}_x}
\int_{-T}^T\log\frac{c(U_x,\omega(s^-))} {c( U^{-1}_x,
U_x\omega(s^-))}dN^{U_x}_s(\omega)\\
&+& \sum_{\begin{array}{c}A\subset\Lambda_n\setminus
\Lambda^{\sharp}_n\\{\mbox{\tt diam}}(A)\leq r
\end{array}}
\sum_{\sigma\in S^A}\int^{T}_{-T} \log\frac{k^{(n)}_A(\omega(s^-),
\omega(s^-)^{A,\sigma})}{k^{(n)}_A(\omega(s^-)^{A,\sigma},\omega(s^-))}
dN^{A,\sigma,n}_s(\omega)
\end{eqnarray*}
The first integral is really a sum over all the times when the trajectory
makes a jump from the action of one of the $U_x$; the second integral is a
sum over all times when a configuration $\sigma$ is replacing
$\omega(s^-)$ in a set $A$ on the boundary. In order to further clarify
this formula, let us first look at trajectories where no boundary
transitions take place (or, what amounts to the same, take $k\equiv 0$
for the moment).  Then, we only keep the first term, that is just
(\ref{vrg}), in case $\pi=id$:
\[
\sum_{x\in\Lambda^\sharp_n} \sum_{U_x\in\mathcal{P}_x}
\int_{-T}^T\log\frac{c(U_x,\omega(s^-))} {c( U^{-1}_x,
U_x\omega(s^-))}dN^{U_x}_s
\]
But if we insert the detailed balance condition (\ref{detbal1}), the
above expression telescopes to
\[
H(\omega(-T))-H(\omega(T))
\]
and the mean entropy production is zero by stationarity.\\
Turning to the general case we let $\{s_i\}_{i=1}^{q}$ be the set of
times at which boundary transitions occur in the sets $A_i,i=1,\ldots,q,$
for the trajectory $\omega$. These are random but we fix them as $-T\leq
s_1<s_2<..<s_{q}\leq T$. The important thing to realize now is that while
the perfect telescoping of above is broken at each of these times, it can
be restored by adding and subtracting.  More precisely, we have
\begin{eqnarray*}
R_n(\omega) &=& H(\omega(-T))-H(\omega(s_1^-))+
H(\omega(s_1))-H(\omega(s_2^-))+
\ldots  \\
&+& H(\omega(s_q))- H(\omega(T))+ \log
\frac{k^{(n)}_{A_1}(\omega(s_1^-),\omega(s_1))}
{k^{(n)}_{A_1}(\omega(s_1),\omega(s_1^-))} + \\
&+& \log \frac{k^{(n)}_{A_2}(\omega(s^-_2),\omega(s_2))}{k^{(n)}_{A_2}
(\omega(s_2),\omega(s^-_2))} + \ldots + \log
\frac{k^{(n)}_{A_q}(\omega(s^-_q),\omega(s^-_q))}
{k^{(n)}_{A_q}(\omega(s_q, \omega(s^-_q))}
\end{eqnarray*}
But by the absolute convergence of the interaction potential we have
\[
|H(\omega(s^-_i)-H(\omega(s_i))|\leq r C
\]
for some constant $C$, since $\omega(s_i^-)$ and $\omega(s_i)$ only
differ in the set $A_i$. Therefore the telescoping of the terms involving
energy differences can be restored upon inserting $q$ terms of order
unity.\\  As for the other terms, we have assumed uniform boundedness so
that we get
\[
|R_n(\omega)|\leq q (r C+\log\frac{M}{\epsilon})
\]
where $M$ and $\epsilon$ are constant upper and lower bounds for the
transition rates $k^{(n)}$.  As the expectation of $q=q(\omega)$ under
$\Ebar_{\rho_n}$ is proportional to $T|\partial\Lambda_n|$, the
proposition is proved.\QED

\noindent {\bf Acknowledgment:}  C.M. thanks Senya Shlosman for some very
useful discussions.


\begin{thebibliography}{99}
\bibitem{G}
Gallavotti G., {\em Breakdown and regeneration of time reversal symmetry
in nonequilibrium Statistical Mechanics\/}  Physica D {\bf 112} 250--257
(1998)

\bibitem{Ku}
K\"unsch, H., {\em Non reversible stationary measures for infinite
interacting particle systems\/} Z. Wahrsch. Verw. Gebiete {\bf 66}, 407
(1984).

\bibitem{M}
Maes C., {\em The Fluctuation Theorem as a Gibbs Property\/}, J. Stat.
Phys. {\bf 95}, 367-392 (1999).
\bibitem{mc}
Maes, C., {\em Statistical mechanics of entropy production: Gibbsian
hypothesis and local fluctuations\/}, preprint from cond-mat/0106464.

\bibitem{MRV}
Maes, C., Redig, F. and Verschuere, M., {\em Entropy Production for
Interacting Particle Systems\/} Markov Proc. Rel. Fields {\bf 7}, 119--134 (2001).

\bibitem{MRVM}
Maes C., Redig F., Van Moffaert A., {\em On the definition of entropy
production via examples} J. Math. Phys. {\bf 41}, 1528--1554 (2000).
\bibitem{MR}
Maes C., Redig F., {\em Positivity of entropy production} J. Stat. Phys.
{\bf 101}, 3--16 (2000).
\bibitem{N} Nakazawa, H., {\em On the Lattice Thermal Conduction} Suppl.
Prog. Theor. Phys., {\bf 45}, 231--262 (1970)
\bibitem{LIGG}
Liggett T.~M., {\em Interacting particle systems} Springer-Verlag, New
York, Heidelberg, Berlin (1985)
\bibitem{LS}
Lebowitz J.~L., Spohn H., {\em A Gallavotti-Cohen type symmetry in the
large deviation functional for stochastic dynamics} J. Stat. Phys. {\bf
95}, 333--365 (1999)

\bibitem{Q}
Qian M.~P., Qian M., Qian C., {\em Circulations of markov chains with
continuous time and probability interpretation of some determinants} Sci.
Sinica {\bf 27}, 470-481. (1984)
\bibitem{QQ}
Qian M.~P., Qian M., {\em The entropy production and reversibility of
Markov processes} Proceedings of the first world congress Bernoulli soc.
1988, 307-316
\bibitem{RLL} Rieder, Z., Lebowitz, J.L. and Lieb, E., {\em Properties of
a Harmonic Crystal in a Stationary Nonequilibrium State} J. Math. Phys.
{\bf 8}, 1073--1078 (1967)

\bibitem{S}
Schnakenberg J., {\em Network theory of behavior of master equation
systems} Rev. Mod. Phys. {\bf 48}, 4, 571-585. (1976)
\end{thebibliography}
\end{document}